\documentclass[10pt,conference]{IEEEtran} 
\usepackage{graphicx} 
\usepackage{amsmath}  
\usepackage{cite}     
\usepackage{url}      
\usepackage{color}

\title{Exploring Structural Complexity in Normative RAG with Graph-based approaches: A case study on the ETSI  Standards}

\author{
    \IEEEauthorblockN{
        Aiman Al Masoud\IEEEauthorrefmark{1}, 
        Marco Arazzi\IEEEauthorrefmark{1}, 
        Simone Germani\IEEEauthorrefmark{2}, and 
        Antonino Nocera\IEEEauthorrefmark{1}
    }
    \IEEEauthorblockA{
        \IEEEauthorrefmark{1}Department of Electrical, Computer and Biomedical Engineering, University of Pavia, Pavia, Italy\\
        Email: \{aiman.almasoud01, marco.arazzi01\}@universitadipavia.it, antonino.nocera@unipv.it
    }
    \IEEEauthorblockA{
        \IEEEauthorrefmark{2}Comitato Elettrotecnico Italiano (CEI), Milano, Italy\\
        Email: simone.germani@ceinorme.it
    }
}

\definecolor{light-gray}{gray}{0.85}

\begin{document}

\maketitle

\begin{abstract}
Industrial standards and normative documents exhibit intricate hierarchical structures, domain-specific lexicons, and extensive cross-referential dependencies, which make them little adequate to be directly processed by pre-trained generative-AI models, such as Large Language Models (LLMs). While Retrieval-Augmented Generation (RAG) provides a computationally efficient alternative to LLM fine-tuning, standard ``vanilla" vector-based retrieval may fail to capture the latent structural and relational features intrinsic in normative documents. With the objective of shedding light on the most promising technique for building high-performance RAG solutions for normative, standards, and regulatory documents, this paper investigates the efficacy of Graph RAG architectures in this complex domain. Graph RAG technology represents information as nodes interconnected through edges, thus moving from simple semantic similarity toward a more robust, relation-aware retrieval mechanism.

Despite the promise of graph-based techniques, there is currently a lack of empirical evidence as to which is the optimal indexing strategy for technical standards. Therefore, to help solve this knowledge gap, we propose a specialized RAG methodology tailored to the unique structure and lexical characteristics of standards and regulatory documents. Moreover, to keep our investigation grounded, we focus on well-known public standards, such as the ETSI EN 301 489 series. We evaluated several lightweight and low-latency strategies designed to embed document structure directly into the retrieval workflow. The considered approaches are rigorously tested against a custom synthesized Q\&A dataset, facilitating a quantitative performance analysis. Our experimental results demonstrate that the incorporation of structural information and relational dependencies can enhance, at least to some extent, retrieval performance and context relevance. Therefore, RAG and, more specifically, the proposed graph-empowered variant can provide a scalable framework for automated normative and standards elaboration.
\end{abstract}

\section{Introduction}

Retrieval Augmented Generation (RAG, for short) is a family of techniques that allow an existing Large Language Model (LLM) to work with fresh data that was not included in its train set (i.e. new and/or private documents).

In fact, transformer-based Large Language Models (LLMs) demonstrate strong language generation and reasoning capabilities but often lack expertise in highly specialized domains. Training or even fine-tuning on custom data requires substantial computational resources, limiting accessibility for many institutions and industries. Retrieval-Augmented Generation (RAG) \cite{gao2024retrievalaugmentedgenerationlargelanguage} addresses these concerns by enriching a pre-trained LLM context with few-shot examples from the target domain, reducing training costs, improving modularity of indexed data, and enhancing traceability.

Generic RAG pipelines are rapidly being adopted in industries. According to the most basic approach for building a RAG solution (or ``vanilla RAG"), document chunks are indexed through a flat list of text embeddings (high-dimensional vector representations of the chunks) and stored in a vector database \cite{gao2024retrievalaugmentedgenerationlargelanguage}. The retrieval is then carried out by comparing the embedding of the user query with existing chunks through a similarity metric, such as cosine similarity.

Despite the success of RAG, standards, regulations, and other technical documents continue to pose a challenge to the approach, as they are highly cross-referential, semi-structured, and make use of specialized formal vocabulary, strongly dependent on the specific domain. 
Furthermore, being technical documents, standards exhibit a highly regular structure, both at: (i) the lexicon level, and (ii) the macro-organizational level.
From the lexicon point of view, these documents tend to use specialized jargon and precise keywords that dense text embeddings (typically used in RAG), which can induce the retrieval mechanism to overlook them in favor of fuzzier semantic similarity. Moreover, standards cover a large variety of domains, each of which uses its own specialized vocabulary that only partially overlaps with the others, which determines a high potential for semantic gaps and misinterpretation across different domains.
At the macro-level, standards are hierarchically organized in sections and sub-sections, making parent-child (and sibling-to-sibling) dynamics between chunks a relevant factor to consider when building a normative content retriever.

In light of the special characteristics mentioned above, one can argue whether common (vanilla) RAG solutions can be adequate to properly operate on standards or whether more advanced strategies and indexing techniques are required.

Graph RAG \cite{han2025retrievalaugmentedgenerationgraphsgraphrag} is a promising approach to dealing with highly-interlinked data. A graph is a data structure composed of nodes, where each pair of nodes may be connected together by an arc. Among their many applications, graphs have recently been employed to organize knowledge in RAG pipelines.

Graph-based retrievers are largely domain-specific: different domains may require different retrieval assumptions (e.g., homophily vs heterophily). Accordingly, Graph RAGs can be classified according to the domain to which they are applied, such as document/citation graphs, knowledge graphs, social networks, and so forth.

Another common classification for Graph RAG solutions is proposed in \cite{zhang2025surveygraphretrievalaugmentedgeneration}. Here, existing approaches are categorized as: (i) index-based (such as \cite{sarthi2024raptor}), (ii) knowledge-based (such as \cite{edge2024local}, \cite{guo-etal-2025-lightrag}), and (iii) hybrid (such as \cite{wu2024medicalgraphragsafe}). In particular, index-based approaches treat the graph solely as an index to retrieve units of information, whereas knowledge-based approaches also use the graph for reasoning and answer generation. Finally, hybrid approaches fall somewhere in the middle of the spectrum.

More advanced (and expensive) approaches to RAG involve agentic loops \cite{singh2025agentic} where LLMs iteratively reason about the completeness and quality of the answer, deciding what tools to call to obtain more information.

Although regulatory documents and standards differ significantly in legal status and purpose, these two types of publication share many similarities and often refer to each other. Several studies \cite{rayo2025hybridapproachinformationretrieval,umar-etal-2025-enhancing} have focused on designing optimized RAG pipelines for regulatory documents. However, only very little effort \cite{Colombo2025} has been devoted to the application of graph-based techniques to improve information retrieval mechanisms for regulatory and standard documents, with promising results.
As a consequence, a consensus on the optimal indexing and retrieval architecture for these technical documents has yet to be established.

Unlike classical vector-based RAG frameworks, Graph RAG conceptualizes data units as nodes within a relational graph, interconnected by edges. The primary advantage of this topology is its ability to allow for efficient exploration of latent dependencies. By traversing edges, the system can retrieve contextual information that may lack direct semantic similarity to the query but is logically linked through bibliographic or structural associations. Ultimately, this leads to effectively capture the heuristic that ``document A requires the consultation of document B''.

To validate this intuition, in this paper, we conduct a systematic analysis and propose a specialized RAG methodology tailored to the unique requirements of standards.
In fact, despite the recent studies mentioned above, which have started to shed light on the potential of structured retrieval, there remains a lack of empirical evidence regarding the comparative performance of lightweight versus graph-heavy approaches within highly regulated technical frameworks.

We evaluate multiple low-latency strategies to embed document structure into retrieval workflows. To validate these approaches, we conducted an experimental comparison on a custom-synthesized benchmark Q\&A dataset generated from a specific subset of the ETSI EN 301 489 series, facilitating a quantitative analysis of retrieval efficacy.

\section{Methodology}

As discussed earlier, most existing efforts to build RAG systems for the regulatory domain adopt a standard, vanilla RAG implementation. 
In line with a few recent studies, we argue that this approach neglects the key characteristics of regulatory documents, which, if properly leveraged, can significantly improve RAG performance.
Since documents in this domain are heavily cross-referenced, rich in citations, hyperlinks, and direct references to other texts, it is natural to model the information they contain as a graph.

\begin{figure}
    \centering
    \includegraphics[width=0.3\textwidth]{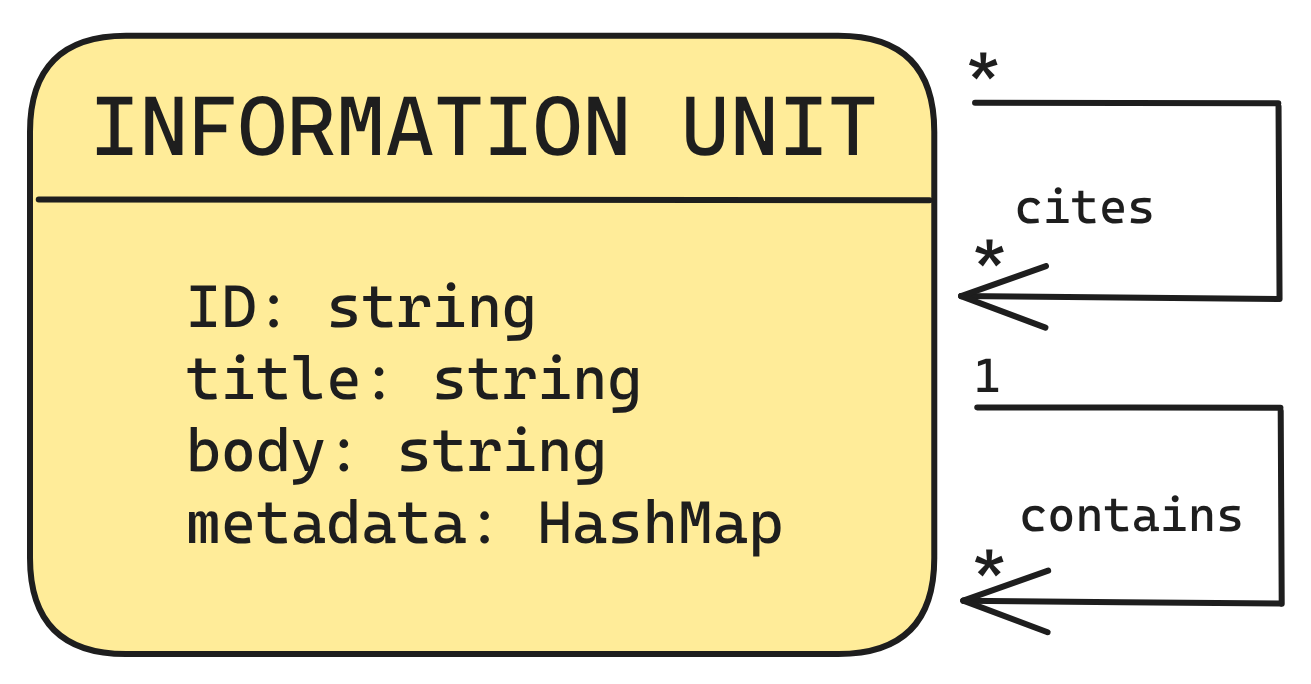}
    \caption{Information Model.}
    \label{fig:info_model}
\end{figure}

In particular, we consider a simple and general Information Model $IG$ for our document graph, dictated by our need to extract the normative content from PDF files each with a slightly different structure (as detailed in Section \ref{sec:graph_construction}); we envision that this simple Information Model can be extended/swapped in the future, when dealing with normative content that is already available in a machine-readable format, such as is foreseen by the IEC Smart Standards framework \cite{iecSmartStandards2024}.

The graph we build consists of homogeneous nodes that we call Information Units $u^i_j \in IU$ (or “InfoUnits”), where $IU$ denotes the complete set of Information Units that compose the graph and $u^i_j$ is part of document $D_i$.
Thus, $IG = (IU, E)$, where the Information Units $IU$ correspond to the nodes of the graph $IG$, and $E$ is the set of edges linking them.

Figure \ref{fig:info_model} presents a high-level depiction of the Information Model $IG$.
Since we are using a property graph metamodel, we envision the InfoUnit as having some intrinsic (or non-relational) properties,

\begin{equation}
u^i_j=\{t,b\}, 
\end{equation}

where $t$ is the title, $b$ is the body of the chunk.

An InfoUnit can take part in two fundamental kinds of relations as follow:

\begin{gather}
    E = P \cup C \\
    P \subseteq IU \times IU, \ C \subseteq IU \times IU
\end{gather}

where $P$ represents parthood relations (document-section or section-subsection), and $C$ represents citation relations between InfoUnits and referenced documents. We assume that each InfoUnit has at most one parent under parthood,

\begin{equation}
\forall u^i_j \in IU,\quad \left| \{ u^i_y \in IU \mid (u^i_j,u^i_y) \in P \} \right| \leq 1,
\end{equation}

with parentless nodes designated as \emph{top-level} InfoUnits.

Ancestry is defined as the transitive closure $P^{+}$ of the parthood relation. An InfoUnit $u^i_j$ is an ancestor of $u^i_y$ if $(u^i_j,u^i_y) \in P^{+}$, and two InfoUnits are considered part of the same publication if they share a common ancestor:

\begin{equation}
\exists u^i_x \in IU:\; (u^i_x,u^i_j) \in P^{+} \wedge (u^i_x,u^i_y) \in P^{+}.
\end{equation}

On the other hand, we assume that citation is a many-to-many relationship; any InfoUnit can cite one or more InfoUnits.

\begin{equation}
\forall u^i_j \in IU,\quad \exists\, C_{u^i_j} \subseteq IU \;\text{s.t.}\; (u^i_j,u^i_y) \in C \;\; \forall u^i_y \in C_{u^i_j}.
\end{equation}

\label{sec:graph_construction}
\subsection{Graph Construction}

\begin{figure}
    \centering
    \includegraphics[width=0.5\textwidth]{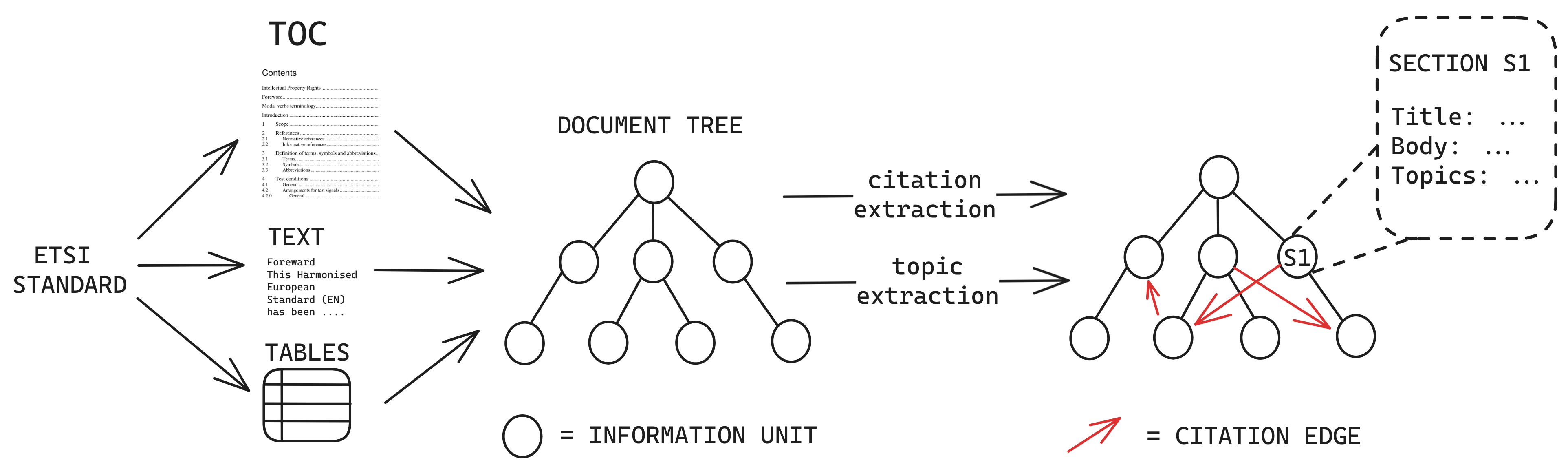}
    \caption{Graph construction.}
    \label{fig:graph_construction}
\end{figure}

Figure \ref{fig:graph_construction} presents a simplified overview of the graph construction pipeline.

We assume that each document $D_i$ has a Table of Contents (ToC), which can be located automatically from its structure. Having located the ToC, we process it to extract the section titles as a sequence of strings, preserving their original order.

After the aforementioned operations, we obtain a flat list of InfoUnits (sections) with intrinsic properties (title and body) but still without any relations.

To obtain the parthood relations, we compare each InfoUnit $u^i_j$ with each subsequent one $u^i_z$, checking if the section code in the title of $u^i_z$ is a valid extension of the section code of $u^i_j$. We assume that the section code is a string of dot-separated characters (e.g. ``1.2.3" or ``A.1.2.3"); a node $u^i_z$ with section code $C_{u^i_v}$ is part of a node $u^i_j$ with section code $C_{u^i_j}$ if $C_{u^i_j}$ is a prefix of $C_{u^i_v}$ and $C_{u^i_v}$ is has exactly one more alphanumerical character than $C_{u^i_j}$.

If a InfoUnit $u^i_j$ exceeds a given length, it can be split into smaller InfoUnits, making the latter parts/children of the former. Tabular sections are exempted from being chunked.

Some InfoUnits can naturally have an empty body, such as when a section directly begins with a title of a sub-section; or when an InfoUnit represents a top-level document.

According to our Information Model, any InfoUnit $u^i_j$ can ``cite" (or ``refer to") any other InfoUnit $u^k_y$ in the Corpus. These references $R$ are categorized as:

\begin{equation}
\text{Type}(R) = 
\begin{cases} 
\text{internal} & \text{if } i = k \\ 
\text{external} & \text{if } i \neq k 
\end{cases}
\end{equation}

When the body of $u^i_j$ only mentions a section (without specifying a standard), we assume that it is an internal reference by default.

We implement a two-stage solution to extract this relational information: (i) parsing the textual mentions within the body, and (ii) resolving each mention to at most one ``referent" InfoUnit. The resolution logic follows the heuristic:

\begin{equation}
\scriptsize
\text{Resolve}(mention) = 
\begin{cases} 
u^i_{section} & \text{if mention contains \{section ID\} only} \\ 
D_k & \text{if mention contains \{doc name\} only} \\
u^k_{section} & \text{if mention contains \{doc name, section ID\}}
\end{cases}
\end{equation}

To resolve a reference, we compare the ``parent" part of the reference to every title/ID in the Corpus; and, assuming we find a matching document $D_k$, we compare the ``child" part of the reference to the title/ID of every child of $D_k$. In practice, to find the parent, we only have to iterate over the top-level documents.

\subsection{Retrieval Pipeline}

\begin{figure}
    \centering
    \includegraphics[width=0.4\textwidth]{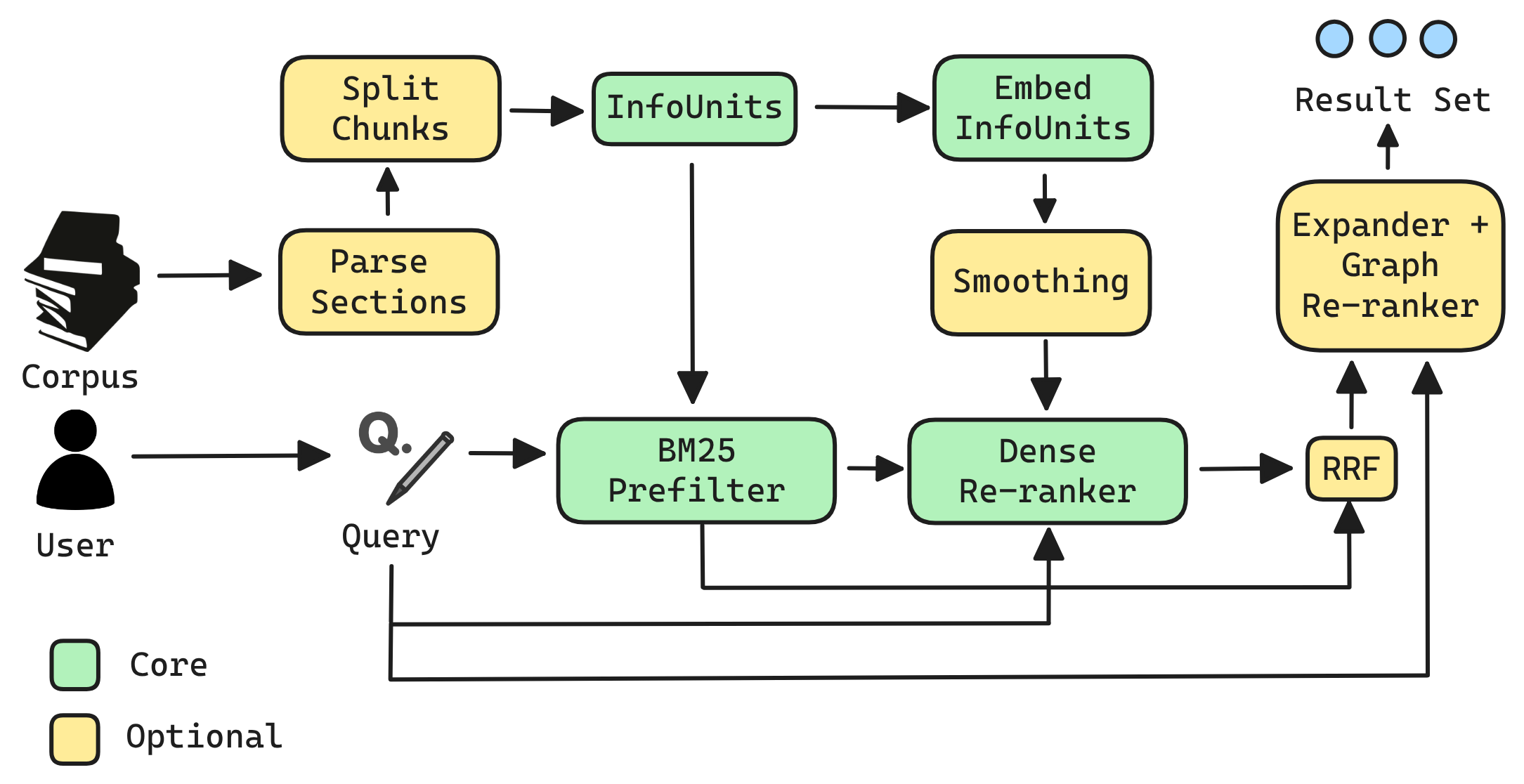}
    \caption{Modular Retrieval Pipeline.}
    \label{fig:modular_retrieval_pipeline}
\end{figure}

To test the effectiveness of the various retrieval-related modules, we build them into a unified pipeline; a high-level overview of the full pipeline is shown in Figure \ref{fig:modular_retrieval_pipeline}. In particular we define seven modules: Parse Sections, Split Chunks, Embed InfoUnits, Smoothing, BM25 Prefilter, Dense Re-ranker, Reciprocal Rank Fusion (RRF), and Expander \& Graph Re-ranker modules.

\label{sec:parse_sections_module}
\subsubsection{Parse Sections}

This is the graph construction module detailed in Section \ref{sec:graph_construction}, it takes the  documents as input $\{D_1,\dots,D_n\}$, and produces a graph of InfoUnits $IG$ as output.

\subsubsection{Split Chunks}

This is a generic chunking component that splits a document $D_i$ into uniform chunks without regard for its official structure. We note that this module can alternatively be combined with the \ref{sec:parse_sections_module} module to split document sections that exceed a certain threshold, in which case the new chunks become children of the original big section.

\label{sec:embed_info_units}
\subsubsection{Embed InfoUnits}

This module uses a small text embedding model (IBM Granite \cite{graniteEmbedding}) $EM(\cdot)$ to compute a dense text embedding representation for each of the InfoUnits obtained by segmenting the corpus as follows:
\begin{equation}
    em^i_j = EM(u^i_j)
\end{equation}
This is done for the benefit of the query-time Dense Re-ranker module (Section \ref{sec:dense_reranker}).

\subsubsection{Smoothing}

This module applies a simple smoothing technique to the embeddings of the nodes in the graph; in particular: it replaces the embedding of each node by the weighted average of: (i) its original value, and (ii) the mean of the values of its neighbors in the graph:


\begin{equation}
em^{i}_{j} \leftarrow \alpha em^{i}_{j} + (1-\alpha) \Sigma_{n,m \in N(u^i_j)} e^{m}_{n}
\end{equation}

Where $\alpha \in (0, 1]$, $em^i_j$ is the embedding of vertex $u^i_j$, and $N(u^i_j)$ is the set of neighbors of $u^i_j$: its children, its parent, its siblings, and the outgoing and incoming citations.

Smoothing has the effect of making the embeddings of connected nodes more similar to each other, irrespective of their intrinsic content; in the form presented above, it is a lightweight alternative to computing graph embeddings for the nodes, such as by using GNNs (Graph Neural Networks).

\label{sec:bm25_prefilter}
\subsubsection{BM25 Prefilter}

The BM25 Prefilter constitutes the initial query-time component of our pipeline, ensuring computational efficiency across large-scale document collections. We utilize Okapi BM25 \cite{robertson1995okapi}, a refinement of the Term Frequency-Inverse Document Frequency (TF-IDF) weighting scheme. This model represents each document $D_i$
as a sparse vector based on the term distribution within the corpus.

Compared to dense embedding-based retrievers, the performance characteristics of BM25 are governed by the following trade-offs:

\begin{equation}
\scriptsize
\text{Performance}(BM25) =
\begin{cases}
\text{High Precision} & \text{if } q \text{ contains exact keywords} \\
\text{Low Latency} & \text{if } |\mathcal{C}| \gg 0 \text{ (large-scale collections)} \\
\text{Limited Recall} & \text{if } q \text{ requires synonymy}
\end{cases}
\end{equation}

While the model excels at lexical matching, it lacks the mechanisms to measure fine-grained semantic relevance or recognize linguistic variations. Consequently, we employ BM25 as a high-recall, low-latency filter to prune the search space before applying more computationally intensive semantic analysis.

\label{sec:dense_reranker}
\subsubsection{Dense Re-ranker}

The dense re-ranker ranks the set of candidates obtained by \ref{sec:bm25_prefilter} based on their semantic similarity to the query, further limiting the set.

\label{sec:rrf}
\subsubsection{Reciprocal Rank Fusion (RRF)}

Reciprocal Rank Fusion (RRF) \cite{10.1145/1571941.1572114} is a technique to combine the outputs of several re-rankers that produce incomparable scores into a single collective ranking:

\begin{equation}
    \mathrm{RRF}(d) = \sum_{r \in Rk} \frac{1}{k + r(d)}
\end{equation}

Where $Rk$ is the set of rankers, $r(d)$ is the score of document $d$ according to ranker $r$, and $k$ is a constant (typically $ \approx 60$). The documents are then sorted by their $\mathrm{RRF}(d)$ score in descending order.

In our case, we use RRF to combine the ranking produced by the dense re-ranker (Section \ref{sec:dense_reranker}) with the one produced by the BM25 prefilter (Section \ref{sec:bm25_prefilter}).

Combining dense and sparse methods is an insight from prior work in Regulatory Information Retrieval \cite{rayo2025hybridapproachinformationretrieval}; the idea being that dense embeddings match the general meaning and intent of a query to the chunks, and keyword-based retrieval methods are beneficial due to the prevalence of precise technical jargon and explicit references within the body of normative documents.

\label{sec:graph_expander}
\subsubsection{Expander \& Graph Re-ranker}

The expander module receives the nodes retrieved by the previous modules and treats them as a set of seeds, expanding each seed by retrieving its neighbors from the graph. 

This module optionally applies a re-ranking strategy to the neighbors, by computing a score that takes into account: (i) a neighbor's connection to each seed element, weighted by the similarity of the seed to the query; (ii) a neighbor's own similarity to the query; and (iii) a ``regularization" component inversely proportional to the degree of the neighbor, to penalize hubs (i.e. nodes that are connected to many other nodes irrespective of the relevance to the query at hand).

\begin{equation}
\begin{aligned}
\text{score}(n) =\;& 
\alpha \sum_{s \in S} \mathbf{1}\{ n \in N(s)\} 
\operatorname{cosim}(em^{s}, em^{q}) \\
&+ \beta \operatorname{cosim}(em^{n}, em^{q}) \\
&- \gamma \log(1 + \operatorname{degree}(n))
\end{aligned}
\end{equation}

Where $S$ is the set of seed nodes retrieved for a given query $q$.

\section{Evaluation}

\subsection{Synthetic Q\&A Pairs}

\begin{figure}
    \centering
    \includegraphics[width=0.2\textwidth]{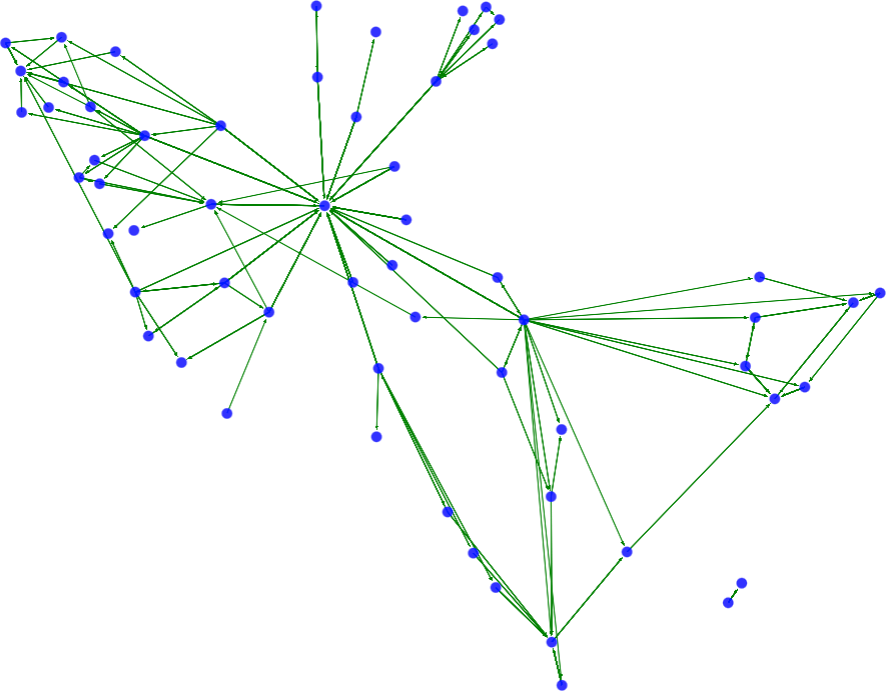}
    \caption{Citations graph of top-level InfoUnits in the corpus.}
    \label{fig:high_level_topology_corpus}
\end{figure}

As aforementioned, we decided to test our method on a batch of ($\approx$ 50) ETSI standards from the ETSI EN 301 489-X series, totalling about 3000 pages (including ToCs and other non-content pages).
As figure \ref{fig:high_level_topology_corpus} shows, the high-level topology of citations in the corpus is mostly connected.

Since we needed a custom Q\&A dataset pertaining to the ETSI standards mentioned above, we chose to generate the Q\&A pairs automatically from the text, using a simple pipeline.

To avoid coupling our evaluation framework to our graph construction framework (Section \ref{sec:graph_construction}), we do not re-cycle any of our graph-construction logic when crafting synthetic Q\&A pairs, opting instead to go back to the original standard documents (in PDF format), and to extract the text from each document as a flat string.

We then create chunks (with a maximum size in tokens each) by splitting these strings. For traceability, we preserve the relation between a chunk and its parent document.

We filter the chunks according to the following heuristics: (i) a chunk must have a mininum number of ``words", where a word is defined as a space-delimited string with a mininum number of characters, we do this to avoid chunks that contain excessive punctuation devoid of any meaning; and: (ii) a chunk must contain a minimum number of modal verbs, as a proxy for its normativity (which in turn makes for interesting Q\&A pairs).

We then sample N chunks at random, and for each of them we generate a Q\&A pair; this is done through a prompt that instructs an LLM (Gemini 2.5-flash \cite{comanici2025gemini25pushingfrontier}) to generate a question that can be answered by the provided context and then answer it concisely. In addition to that, the LLM is instructed to faithfully quote the specific, most relevant, spans of provided context that were used to answer the question (which we call ``witness strings" or ``fine-grained golden chunks").

During the evaluation of an answer to a Q\&A pair, the witness strings are used as golden chunks to be compared to the chunks returned by the retriever. Due to the fact that our evaluation framework is decoupled from our index construction logic, the golden chunks rarely, if ever, match the retrieved chunks perfectly. To compensate for this mismatch, we stipulate that a retrieved chunk match a golden chunk if at least 75\% of the golden chunk is contained as a substring in the retrieved chunk.

We generate 1000 Q\&A pairs from our corpus using the method outlined above. To avoid penalizing any retrieval strategy, we filter out the questions whose golden chunks do not appear in all of the of the constructed indexes' full texts; this can happen for a number of reasons: such as the LLM misquoting the original text (especially when rare unicode characters are involved), or the index we built not including the text because it was not included in the parsed sections. After filtering, we end up with more than 800 valid Q\&A pairs.

\subsection{Metrics}

To evaluate our results, we make use of the following metrics:

\subsubsection{Recall at K (R@K)} is simply the percentage of golden chunks that were actually present among the $K$ chunks returned by the retriever: 

\begin{equation}
    R@K = \frac{|C_R(K) \cap C_G|}{|C_G|}
\end{equation}

Where $C_G$ is the set of golden chunks, and $C_R(K)$ is the set of $K$ retrieved chunks.

\subsubsection{Average Precision at K (AP@K)} is an average measure of the precision (i.e. ratio of true positives to positives) that only notes the ranks where a relevant item has been encountered:

\begin{equation}
AP@K = \frac{1}{r} \sum_{k=1}^{K} \frac{\sum_{i=1}^{k} \mathbf{1}_{\text{rel}}(i)}{k}\, \mathbf{1}_{\text{rel}}(k)
\end{equation}

\subsubsection{(Mean) Reciprocal Rank (M)RR@K} is a measure of how early the first relevant chunk appears among the results:

\begin{equation}
\mathrm{MRR@K} = \frac{1}{|Q|} \sum_{i=1}^{|Q|} \frac{1}{\operatorname{rank}_i}
\end{equation}

\subsection{Experimental Results}

\begin{figure}
    \centering
    \includegraphics[width=0.5\textwidth]{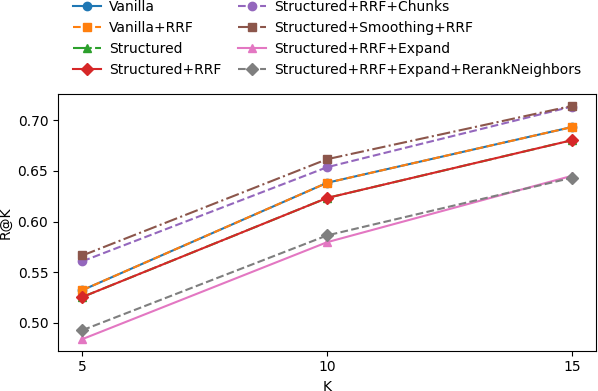}
    \caption{R@K compared across methods.}
    \label{fig:recall8}
\end{figure}

\begin{figure}
    \centering
    \includegraphics[width=0.5\textwidth]{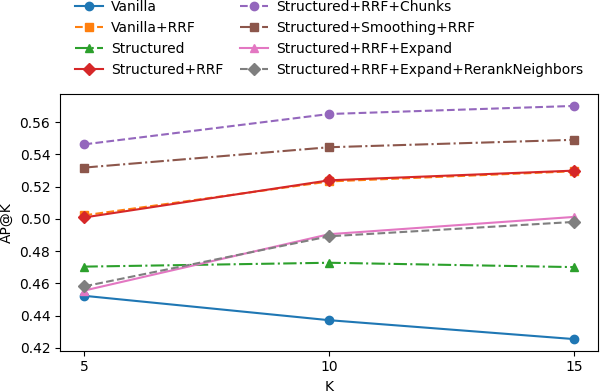}
    \caption{AP@K compared across methods.}
    \label{fig:precision8}
\end{figure}

\begin{figure}
    \centering
    \includegraphics[width=0.5\textwidth]{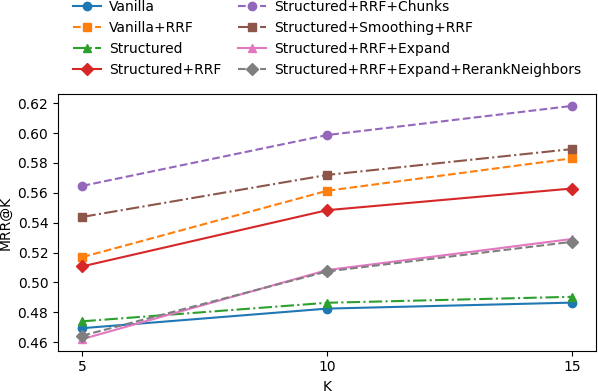}
    \caption{MRR@K compared across methods.}
    \label{fig:mrr8}
\end{figure}

We ran the experiments with the 800+ Q\&A pairs on 8 different configurations of the retriever for 3 different values of K. The variants tagged ``Vanilla" make use of a flatter index constructed through plain chunking (max 300 words per chunk), while the variants tagged ``Structured" take into account the official section hierarchy of the standards. The variant tagged both ``Structured" and ``Chunks" combines the strategies by making use of the official sections, splitting them into sub-sections when they exceed the 300 words limit.

The figures \ref{fig:precision8}, \ref{fig:recall8} and \ref{fig:mrr8} show the results of the experiment in terms of precision, recall and Mean Meciprocal Rank respectively, all aggregated by retriever configuration.

We observe that preserving the structure has a positive effect on precision and MRR, but does not seem to have a statistically significant impact on recall.

Mixing dense and sparse rankings with RRF also has a positive effect on precision and MRR, as expected for the reasons outlined in Section \ref{sec:rrf}; obviously, it does not have any effect on recall.

Neighbor expansion \ref{sec:graph_expander} and re-ranking does not seem to be effective, indicating that more work is needed to determine a neighbor-scoring function that is adequate for the domain.

Smoothing slightly improves recall compared to just preserving the structure and/or using chunks, indicating it as a promising way of embedding relational information into the index for the purpose of improving the completeness of responses.

Overall, the methods we have tested mostly improve the precision of retrieval, and structure-preservation with chunking seems to be the overall best compromise. 

\section{Conclusions}

In this paper, we investigated lightweight, low-latency techniques to enhance RAG effectiveness for standards and normative documents. Our approach investigates the integration of structural information into the indexing mechanism and assesses the performance impact of this choice. Our results demonstrate that the native hierarchical structure and lexical regularities of standards offer a robust framework for improving retrieval precision. We conclude that leveraging these inherent structural assets allows the retrieval pipeline to move beyond simple semantic similarity, better addressing the complex, cross-referential nature of technical corpora.

Although this study can represent a foundation for the research community, future research should evaluate these methods against broader-scoped Q\&A datasets presenting more complex links in responses. In particular, the capability of these approaches to synthesize information from distant passages inside single documents or across disparate documents needs to be further investigated. In fact, in such cases, hybrid solutions could be a high-priority direction to achieve comprehensive automated normative analysis and elaboration.




\bibliographystyle{plain}

\end{document}